\documentclass[journal=jacsat,manuscript=article]{achemso}

\usepackage[version=3]{mhchem} 
\usepackage[T1]{fontenc}
\usepackage[latin1,utf8]{inputenc}
\usepackage{physics}
\usepackage{textcomp}
\usepackage{color}
\usepackage{units}
\usepackage{amssymb}
\usepackage{graphicx}
\usepackage{esint}
\usepackage{bm}
\usepackage{natbib}
\usepackage{ulem}
\usepackage{tabularx}
\usepackage{float}
\usepackage{gensymb}
\usepackage{xcolor}
\usepackage[colorlinks=true, citecolor=black]{hyperref}
\usepackage{lipsum}
\usepackage{xfrac}
\usepackage{caption}
\DeclareCaptionLabelSeparator{vertbar}{$~\bm \vert~$}
\usepackage{sidecap}  
\usepackage{babel}
\usepackage{multirow} 
\usepackage{dcolumn}

\author{Thasneem Aliyar}
\affiliation{Division of Physics and Applied Physics, School of Physical and Mathematical Sciences, Nanyang Technological University, Singapore 637371, Singapore}

\author{Hongyang Ma}
\affiliation{School of Physics, University of New South Wales, Sydney, NSW 2052, Australia}

\author{Radha Krishnan}
\affiliation{Division of Physics and Applied Physics, School of Physical and Mathematical Sciences, Nanyang Technological University, Singapore 637371, Singapore}

\author{Gagandeep Singh}
\affiliation{Division of Physics and Applied Physics, School of Physical and Mathematical Sciences, Nanyang Technological University, Singapore 637371, Singapore}

\author{Bi Qi Chong}
\affiliation{Division of Physics and Applied Physics, School of Physical and Mathematical Sciences, Nanyang Technological University, Singapore 637371, Singapore}

\author{Yitao Wang}
\affiliation{Division of Physics and Applied Physics, School of Physical and Mathematical Sciences, Nanyang Technological University, Singapore 637371, Singapore}

\author{Ivan Verzhbitskiy}
\affiliation{Institute of Materials Research and Engineering (IMRE), Agency for Science, Technology and Research (A*STAR), 138634, Singapore}

\author{Calvin Pei Yu Wong}
\affiliation{Institute of Materials Research and Engineering (IMRE), Agency for Science, Technology and Research (A*STAR), 138634, Singapore}

\author{Kuan Eng Johnson Goh}
\affiliation{Institute of Materials Research and Engineering (IMRE), Agency for Science, Technology and Research (A*STAR), 138634, Singapore}

\alsoaffiliation{Division of Physics and Applied Physics, School of Physical and Mathematical Sciences, Nanyang Technological University, Singapore 637371, Singapore}

\author{Ze Xiang Shen}
\affiliation{Division of Physics and Applied Physics, School of Physical and Mathematical Sciences, Nanyang Technological University, Singapore 637371, Singapore}

\author{Teck Seng Koh}
\affiliation{Division of Physics and Applied Physics, School of Physical and Mathematical Sciences, Nanyang Technological University, Singapore 637371, Singapore}

\author{Rajib Rahman}
\affiliation{School of Physics, University of New South Wales, Sydney, NSW 2052, Australia}

\author{Bent Weber}
\email{b.weber@ntu.edu.sg}
\affiliation{Division of Physics and Applied Physics, School of Physical and Mathematical Sciences, Nanyang Technological University, Singapore 637371, Singapore}
\email{b.weber@ntu.edu.sg}

\title[An \textsf{achemso} demo]
  {Symmetry breaking and spin-orbit coupling for individual vacancy-induced in-gap states in MoS$_2$ monolayers}

\keywords{American Chemical Society, \LaTeX}

\begin{document}

\begin{abstract}
\sloppy
  Spins confined to point defects in atomically-thin semiconductors constitute well-defined atomic-scale quantum systems that are being explored as single photon emitters and spin qubits. Here, we investigate the in-gap electronic structure of individual sulphur vacancies in molybdenum disulphide (MoS$_2$) monolayers using resonant tunneling scanning probe spectroscopy in the Coulomb blockade regime. Spectroscopic mapping of defect wavefunctions reveals an interplay of local symmetry breaking by a charge-state dependent Jahn-Teller lattice distortion that, when combined with strong ($\simeq$100 meV) spin-orbit coupling, leads to a locking of an unpaired spin-$1/2$ magnetic moment to the lattice at low-temperature, susceptible to lattice strain. Our results provide new insights into spin and electronic structure of vacancy induced in-gap states towards their application as electrically and optically addressable quantum systems.
  
  \textbf{Keywords:} TMDC, MoS$ _2$, Coulomb blockade, Scanning tunneling microscopy and spectroscopy (STM/STS), resonant tunneling, atomic-scale quantum systems, in-gap states.
\end{abstract}

\sloppy
Atomic point defects in semiconductors and insulators that confine individual charges and spin constitute well-defined atomic-scale quantum systems \cite{koenraad2011single,voisin2023solid,salfi2014spatially} with applications in quantum computing, communication, sensing, and simulation \cite{wolfowicz2021quantum,atature2018material,chatterjee2021semiconductor,de2021materials}. In traditional 3D semiconducting and insulating platforms, such as silicon \cite{pla2012single,weber2010quantum}, diamond \cite{doherty2013nitrogen} or silicon carbide (SiC) \cite{chatterjee2021semiconductor}, point defects  addressed either optically or electrically, have allowed to realize quantum bits (qubits) \cite{greentree2008diamond,pla2012single,weber2010quantum,maurer2012room} and single photon emitters (SPE) \cite{bathen2021manipulating,aharonovich2016solid}.

More recently, atomically-thin van-der-Waals (vdW) semiconductors and insulators have emerged as candidate platforms for optoelectronics applications \cite{wang2012electronics,manzeli20172d} and quantum technologies \cite{kormanyos2014spin,liu20192d}. In the semiconducting transition metal dichalcogenides (TMDCs), this is owing to their large and direct band gap in the monolayer limit \cite{manzeli20172d,chhowalla2013chemistry}, strong spin-orbit coupling \cite{kormanyos2014spin}, and additional valley degree of freedom, with spin-valley coupling \cite{xu2014spin,krishnan2023spin,schuler2019large}. Point defects in TMDCs have been identified to introduce strongly confined electronic bound states within the bandgap $-$ in-gap states  \cite{hong2015exploring,qiu2013hopping,tsai2022antisite,klein2019site,mitterreiter2021role}$-$ that can inherit these properties \cite{krishnan2023spin}, promising applications as optically and electrically addressable quantum systems. 

\begin{figure}[t]
\begin{center}
\includegraphics[scale=0.479]{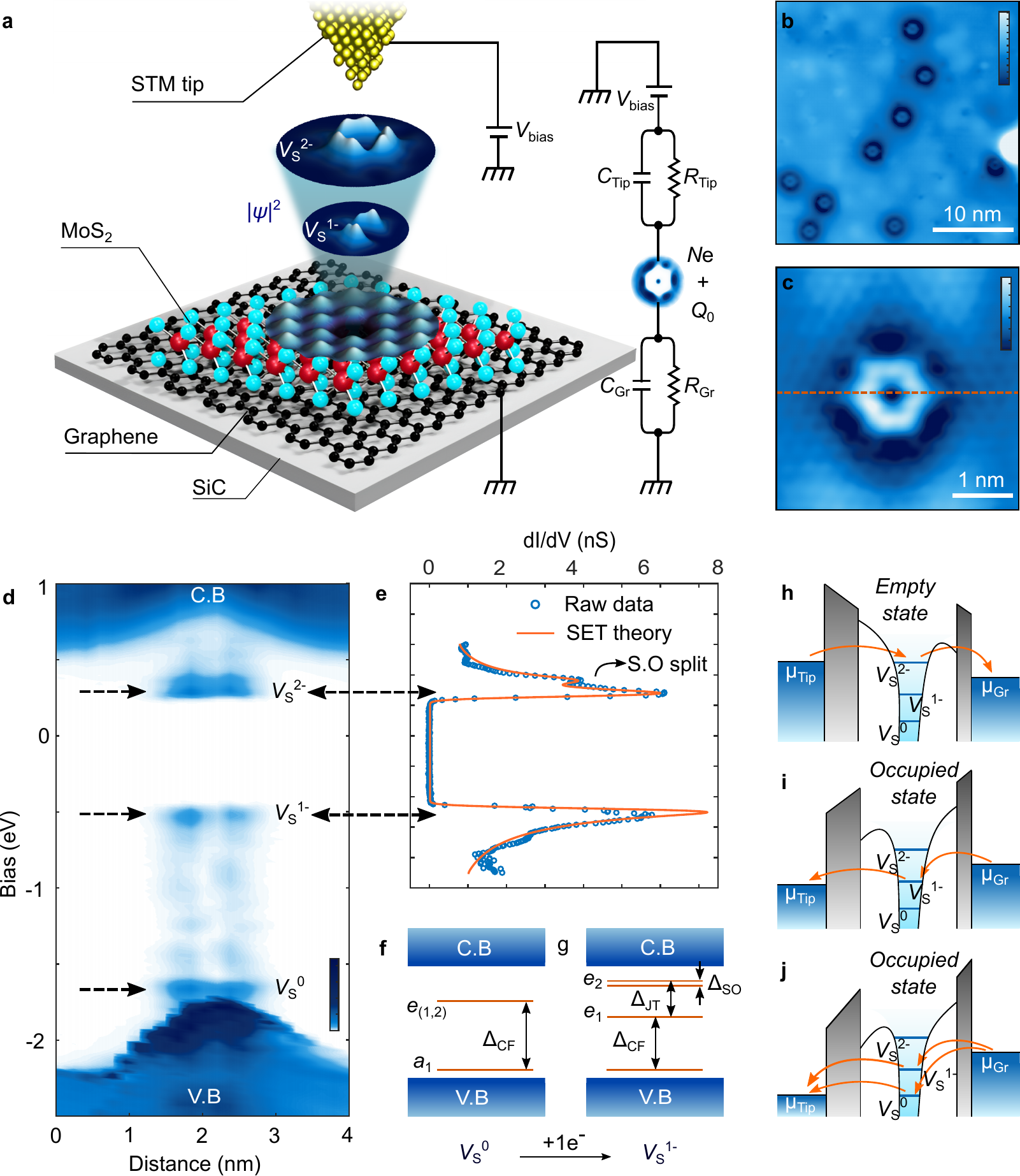}
\captionsetup{labelfont=bf,name={Figure},labelsep=period}
\caption {\small{\textbf{Spatial mapping of the in-gap charge states of individual sulphur vacancy defects. a,} Schematic of the experiment comprised of a tungsten tip positioned above a  MoS$_2$ monolayer on an epitaxial graphene on SiC substrate. The equivalent circuit represents the double barrier structure, formed by the vacuum barrier between tip and defect, and defect and Gr/SiC substrate, respectively. \textbf{b,} Large area topographic scan of the MoS$_2$ monolayer showing several point defects ($V\textsubscript{bias}$ = 650 mV, $I$ = 30 pA). \textbf{c,} Close-up of a single sulphur vacancy (V$_{\rm S}$) defect ($V\textsubscript{bias}$ = 600 mV, $I$ = 30 pA). \textbf{d,} Line spectroscopy across the same S vacancy (lock-in modulation ($V\textsubscript{ac}$=20 mV) along the orange dashed line in \textbf{c}. In-gap states are indicated by dashed black arrows. The pronounced upward band bending observed around the defect indicates a negative charge localized to the defect. The inverted parabola at negative bias indicates a tip-induced change in the equilibrium charge state. \textbf{e,} High resolution STS ($V\textsubscript{ac}$=5mV). The orange line shows a fit to Coulomb blockade theory. Energy level schematic of V$_{\rm S}$ in-gap states for both (\textbf{f}) neutral and (\textbf{g}) charged vacancy. \textbf{h-j,} Schematics, showing tunneling through the various charge states at both positive (\textbf{h}) and negative (\textbf{i,j}) bias. 
\label{fig:fig1}}}.
\end{center}
\end{figure}

Chalcogen vacancies are the most prevalent naturally occurring point defect in TMDCs \cite{noh2014stability,santosh2014impact}. In-gap states due to sulphur (S) vacancies in MoS$_2$ give rise to its ubiquitous $n$-type doping \cite{di2022defects} and can allow to bind excitons in DX$^0$ centres, giving rise to defect-induced \cite{zhu2023room}  or single-photon emission  \cite{klein2019site,mitterreiter2021role}. Spin-valley locking for in-gap states has been detected in a large valley Zeeman effect \cite{wang2020spin} and Zeeman anisotropy \cite{krishnan2023spin}. Spectroscopic signatures of in-gap states \cite{schuler2019large,trainer2022visualization,ponomarev2018hole}, and photon emission from individual vacancy defects \cite{schuler2020electrically} have been captured at the atomic-scale by scanning tunneling microscopy (STM).

In this work, we employ resonant tunneling scanning probe spectroscopy in the Coulomb blockade regime, to probe the in-gap electronic structure of individual chalcogen vacancies (V$\textsubscript{S}$) in MoS$_2$ monolayer at 4.5 K. Spectroscopic mapping of the defect wave functions for two distinct charge states reveals an interplay of local symmetry breaking by a charge-state dependent Jahn-Teller (JT) lattice distortion, combined with strong spin-orbit coupling (SOC). As confirmed by DFT, this leads to a vacancy-bound spin-$1/2$ magnetic moment that is locked to the lattice at low temperature, susceptible to lattice strain as reflected in wavefunction symmetry and orientation. 

Figure~\ref{fig:fig1}a shows a schematic representation of our experiment. A biased tungsten tip is brought into close proximity of a monolayer of mechanically exfoliated natural MoS$_2$, transferred onto a graphene monolayer (see Methods) that was epitaxially grown on silicon carbide (G/SiC), and is connected via a metal clamp to the STM system ground. Topographic STM images (Figure~\ref{fig:fig1}b,c) of the MoS$_2$ monolayer show various point defects, including sulphur vacancies (V$\textsubscript{S}$) and oxygen substituted/passivated sulphur vacancies (O$\textsubscript{S}$) \cite{barja2019identifying}. Among them, unpassivated sulphur vacancies have the clearest topographic signature owing to their negative V$\textsubscript{S}$$^{1-}$ charge state at high doping (Figure~\ref{fig:fig1}b) \cite{tan2020stability,schuler2020electrically}. Each isolated V$\textsubscript{S}$ vacancy defects is vertically separated from the tungsten tip by a vacuum barrier and from the G/SiC substrate by a van-der-Waals gap (Figure~\ref{fig:fig1}b), forming a double-barrier structure (Figure~\ref{fig:fig1}h-j) \cite{likharev1999single} that can be described by an equivalent circuit as shown in Figure~\ref{fig:fig1}a. 

\begin{figure*}[t]
\begin{center}
\includegraphics[scale=0.7]{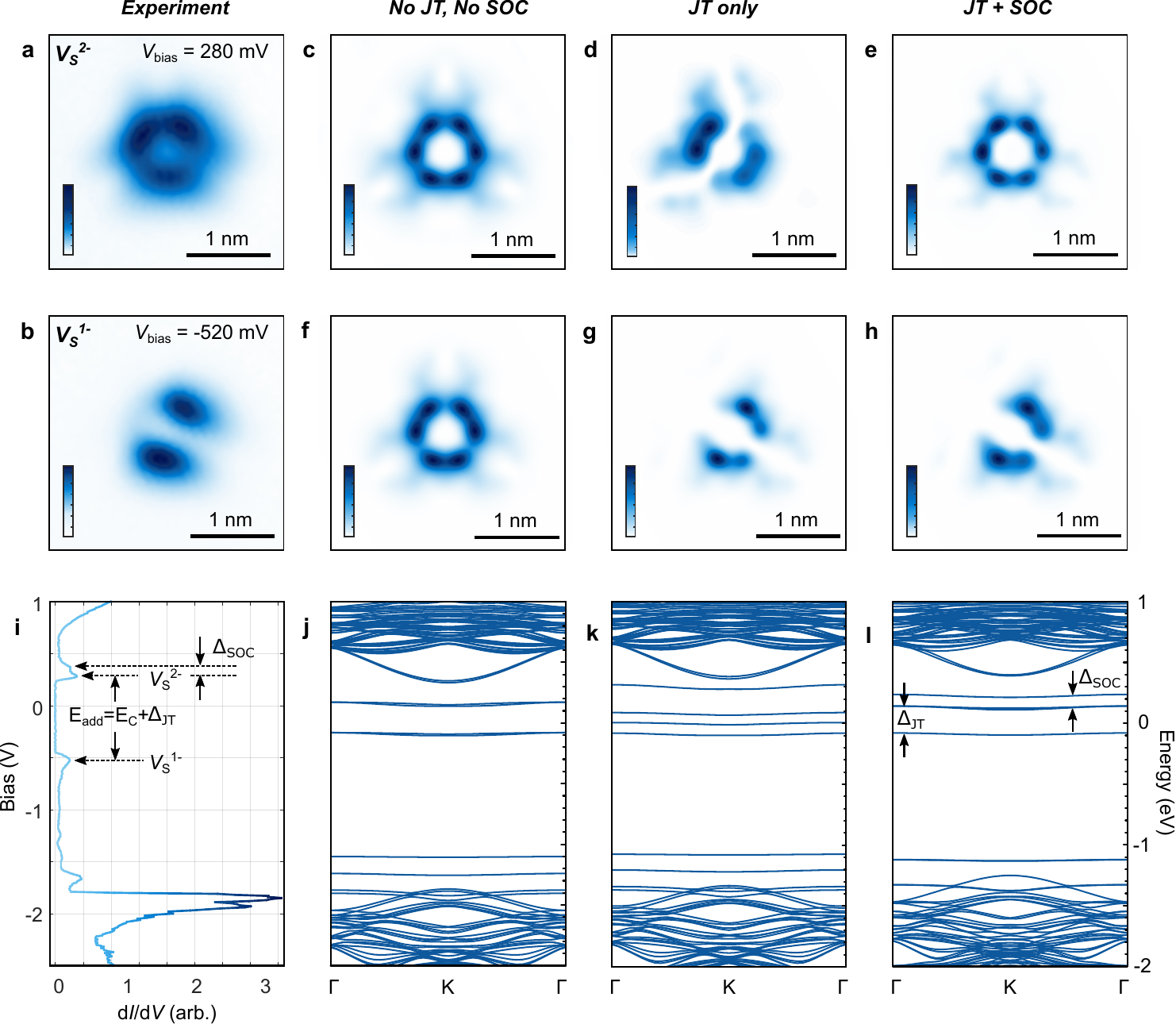}
\captionsetup{labelfont=bf,name={Figure},labelsep=period}
\caption {\small{\textbf{Interplay of a Jahn-Teller lattice distortion with spin-orbit coupling (SOC).} \textbf{a-h,} Comparison of the measured (\textbf{a-b}) and DFT calculated (\textbf{c-h}) defect wavefunctions for in-gap states at $+$280 mV (\textbf{a}) and $-$520 mV (\textbf{b}), confirming that wavefunction symmetry is determined by the presence of both, a Jahn-Teller lattice distortion and spin-orbit coupling. \textbf{i,} STM point spectrum (V$\textsubscript{mod}=$10 mV) showing the presence of an additional excited state owing to a spin-orbit splitting of the V$\textsubscript{S}$$^{2-}$ ($e\textsubscript{2}$) state. \textbf{j-l,} Corresponding DFT band structure of defective MoS$_2$ under the same conditions as in (\textbf{c-h}) above.
\label{fig:fig2}}}.  
\end{center}
\end{figure*}

A series of point spectra across an isolated V$\textsubscript{S}$ defect (dashed line in Figure~\ref{fig:fig1}c) is 
shown in Figure~\ref{fig:fig1}d, in which we can identify three distinct in-gap states (dashed arrows) that are strongly confined to within $\sim$ 2 nm around the defect location. A feature resembling an inverted parabola, centered at the defect, is attributed to tip-induced band bending, and reflects a tip-induced change in the defect's equilibrium charge state \cite{di2022defects,schuler2019large,zhang2021observation}.

We understand the formation of V$\textsubscript{S}$ induced in-gap states as a result of dangling Mo $d$ orbitals around the vacancy defect \cite{santosh2014impact}. In the pristine MoS$_2$ lattice, each Mo atom is coordinated by six chalcogens in a trigonal prismatic ($D\textsubscript{3h}$) configuration  \cite{li2016strong}. Electrostatic repulsion from the surrounding chalcogens (the crystal field) lifts the five-fold degeneracy of the $d$ manifold, with a doubly-occupied $d_{z^{2}}$ orbital being the lowest in energy, followed by degenerate pairs of $d_{xy}$/$d_{x^{2}-y^{2}}$ and $d_{xz}$/$d_{yz}$ orbitals above \cite{li2016strong}. In defect-free MoS$_2$, the Mo atoms assume an oxidation state of Mo$^{4+}$, providing 4 electrons to bonds with the six surrounding chalcogens. The two S atoms per unit cell each assume an oxidation state of S$^{2-}$. Removing a S atom from the lattice to create a vacancy will therefore not only leave three dangling Mo $d$-orbitals but also free up two electrons \cite{lu2018passivating,pandey2016defect}. The dangling Mo $d$ orbitals \cite{vancso2016intrinsic,naik2018substrate} create three in-gap states (Figure~\ref{fig:fig1}f), one state ($a_1$) close to the valence band, and two degenerate states ($e_{1,2}$) near the conduction band   \cite{tan2020stability} as reflected in our data (Figure~\ref{fig:fig1}d). Our own DFT calculations confirm (see Methods), that orbital contributions of the $a_1$ state are dominated by $d_{xz}$ and $d_{yz}$ orbitals, while the $e_{1,2}$ states are dominated by $d_{xy}$ and $d_{x^{2}-y^{2}}$ orbitals (see supplementary information for detail). In a charge-neutral V$\textsubscript{S}$-vacancy, the $a_1$ state remains occupied by the two excess electrons, whereas the $e_{1,2}$ states remain unoccupied. However, the natural MoS$_2$ crystals used in this work are highly doped, as reflected in the position of Fermi level energy ($E=E\textsubscript{F}$) closer to the conduction band edge (Figure~\ref{fig:fig1}d). High doping allows population of the $e_{1,2}$ manifold changing the vacancy charge state to V$\textsubscript{S}$$^{1-}$ \cite{noh2014stability}, as confirmed by the observed upward bending of the conduction band edge in the vicinity of the defect (Figure~\ref{fig:fig1}d).

The presence of this additional electronic charge has a profound effect on the electronic degeneracies and the spin-orbit splitting as it introduces a Jahn-Teller (JT) lattice distortion  -- a simultaneous electronic and geometric symmetry breaking at the defect site \cite{tan2020stability}. In particular, the JT lattice distortion has been theoretically predicted (but not yet experimentally shown) to lift the degeneracy of the two-fold degenerate $d_{xy}$ and $d_{x^{2}-y^{2}}$ orbitals ($e_{1,2}$) with an additional electron occupying the lower ($d_{x^{2}-y^{2}}$) of the two \cite{tan2020stability}. The symmetry breaking also gives rise to an enhancement of the spin-orbit splitting of the $d_{xy}$ and $d_{x^{2}-y^{2}}$ orbitals, \cite{li2016strong} estimated to be $\simeq$100 meV in MoS$_2$ according to our DFT calculations.

For an accurate experimental extraction of the single particle energies, and in particular the spin-orbit splitting, Coulomb charging of the microscopic junction capacitances of the STM needs to be taken into account. For instance, Coulomb charging is clearly reflected in the appearance of a pronounced Coulomb gap at the Fermi energy ($E=E\textsubscript{F}$) with strongly suppressed tunneling conductance ($dI/dV$) over an energy range of $\simeq$800 meV. The gap is bounded by two highly asymmetric conductance peaks, reflecting the V$\textsubscript{S}$$^{0}$--V$\textsubscript{S}$$^{1-}$ and V$\textsubscript{S}$$^{1-}$--V$\textsubscript{S}$$^{2-}$ ground state transitions on the S vacancy defect respectively. A change in the total energy $E\textsubscript{add}$ (the addition energy) of the defect state as a result of adding a single electronic charge equals the sum of single-particle level splitting ($\Delta$) and Coulomb charging energy  $E\textsubscript{C}$ = $\frac{e^2}{2(C\textsubscript{tip}+C\textsubscript{Gr})}$, where $C\textsubscript{tip}$ and $C\textsubscript{Gr}$ are the tip and substrate junction capacitances (Figure~\ref{fig:fig1}a). Their respective contributions can be disentangled from a fit (orange line in Figure~\ref{fig:fig1}e) to Coulomb blockade (CB) theory (see supplementary section II for details) from which we extract a large Coulomb charging energy $E\textsubscript{C}= (211 \pm 2)$~meV and a single particle splitting $\Delta\textsubscript{JT} = (225 \pm 3)$~meV in the V$\textsubscript{S}$$^{1-}$ ($e_1$) state that we identify as the JT energy scale, lifting the degeneracy of the $e_{1,2}$ manifold  \cite{gupta2018two}. Spin-orbit splitting of the V$\textsubscript{S}$$^{2-}$ ($e_2$) state manifests as a low lying excited state at $\Delta\textsubscript{SOC}$ $\sim$84 meV above the V$_{\rm S}^{1-}$--V$_{\rm S}^{2-}$ ground state transitions (Figure~\ref{fig:fig1}e, arrow), in good agreement with our DFT calculations ($\simeq$100 meV) as shown in Figure~\ref{fig:fig2}l. 

To further confirm the orbital character of the in-gap wavefunctions and unravel the respective roles of the Jahn-Teller lattice distortion and SOC, we turn to constant-height conductance mapping to resolve the probability density of the in-gap wavefunction as reflected in the local density of states, $\rho\textsubscript{LDOS}(\textbf{r},E)=\sum_{i}|\psi(\textbf{r})|^2\delta(E_i-E)$. As shown in Figure~\ref{fig:fig2}, we observe a threefold symmetric defect wavefunction in the V$\textsubscript{S}$$^{2-}$ charge state (Figure~\ref{fig:fig2}a)  reflecting the $D\textsubscript{3h}$ symmetry of the MoS$_2$ lattice. The wavefunction of the V$\textsubscript{S}$$^{1-}$ charge state, on the other hand, shows a mirror-symmetric character (Figure~\ref{fig:fig2}b), reflecting the combined effects of symmetry lowering by the JT distortion and SOC. Indeed, our DFT calculations confirm that the respective wavefunctions of the V$\textsubscript{S}$$^{1-}$ and V$\textsubscript{S}$$^{2-}$ charge states can only be simultaneously captured under inclusion of both effects (JT+SOC in Figure~\ref{fig:fig2}e,h), also reflected in the calculated quasiparticle bandstructures (Figure~\ref{fig:fig2}j-l). Only Figure~\ref{fig:fig2}l (JT+SOC) shows two in-gap states straddling the Fermi level ($E=0~$eV), with a single SOC excited state at  $\Delta\textsubscript{SOC}= 100$~meV above. From these single particle band structures, we extract the single particle splitting of the  V$\textsubscript{S}$$^{1-}$ as $\Delta\textsubscript{JT}$ $=$ 222 meV, in excellent agreement with that extracted from our CB fits (225 meV). Fourier analysis of the in-gap probability density states \cite{liu2015observation,yankowitz2015local,salfi2014spatially,simon2009symmetry} further reveals that the in-gap states draw
from the conduction band $Q$ valleys, instead of $K$, even at the monolayer limit (see Figure 3 supplementary information for details), in which the JT distortion causes a shift in valley weights pronounced.

\begin{figure*}[t]
    \centering
    \includegraphics[scale=0.98]{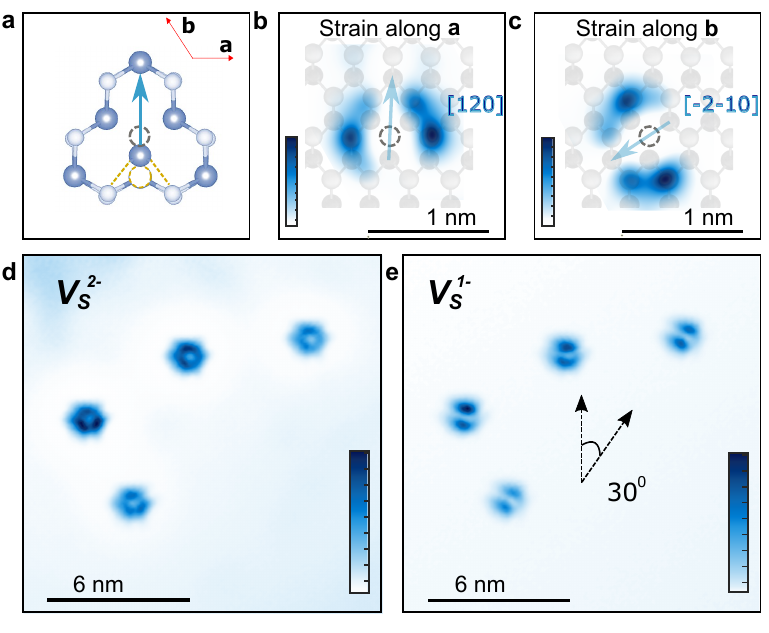}
    \captionsetup{labelfont=bf,name={Figure},labelsep=period}
    \caption{\small \textbf{Strain control of a spin-$1/2$ magnetic moment. a,} The combined effects of Jahn-Teller lattice distortion and SOC couple an unpaired spin-$1/2$ moment (blue arrow) to the lattice, making its orientation susceptible to strain.  
    This is confirmed by DFT calculations shown in \textbf{b,c}, in which uniaxial strain of (1\%) has been applied along two high-symmetry directions of the crystal as indicated. \textbf{d,e}, dI/dV wavefunction mapping of multiple sulphur vacancies in their V$\textsubscript{S}$$^{2-}$ ($+$340 mV) and V$\textsubscript{S}$$^{1-}$ ($-$520 mV) charge states, respectively.}
    \label{fig:fig3}
\end{figure*}

Given that the V$\textsubscript{S}$$^{1-}$ charge state is expected to host an unpaired spin-$1/2$ electron, we calculate the ground-state energy for different possible spin orientations, thus to determine the spin quantization axis with respect to the atomic lattice. We find that the spin magnetic moment is locked into the crystal plane and oriented parallel to the mirror symmetry axis of the V$\textsubscript{S}$$^{1-}$ wavefunction (blue arrow in Figure~\ref{fig:fig3}a). Importantly, this makes the defect wavefunction an indicator of spin polarization, as further confirmed in Figure~\ref{fig:fig3}, showing an area of multiple S-vacancies. While the V$\textsubscript{S}$$^{2-}$ wavefunctions (Figure~\ref{fig:fig3}d) appear identical at each defect site, the V$\textsubscript{S}$$^{1-}$ wavefunctions (Figure~\ref{fig:fig3}e) appear to take on different orientations from site to site, with their mirror axes aligned to the high-symmetry axes of the MoS$_2$ lattice, roughly 30$^{\circ}$ apart. The likely reason for these differences in orientations are local variations in lattice strain, given the importance of the Jahn-Teller lattice distortion (Figure~\ref{fig:fig3}a). 

DFT calculations of the V$\textsubscript{S}$$^{1-}$ wavefunction under application of in-plane uniaxial lattice strain (Figure~\ref{fig:fig3}b,c) confirms that strain preserves the mirror symmetry of the wavefunction, but has the effect of rotating the mirror axis within the plane. Such interplay of lattice strain and a Jahn-Teller induced coupling of  spin and lattice, could thus be an interesting avenue towards realizing straintronic applications, or even strain control of vacancy-induced magnetism \cite{tao2014strain,yun2015strain,salami2016tunable,li2018charge}. The magnetoanisotropy energy (MAE) can be enhanced through application of uniaxial strain along $\bm{a}$ and $\bm{b}$ directions of the crystal and results in an in-plane MAE of $6~\mu$eV along $\bm{a}$ and $9~\mu$eV along the $\bm{b}$ direction, respectively, which may be relevant to applications in atomic-scale spin-based qubits and other single-spin devices operated at cryogenic temperatures $k_{\rm B}T \sim \textrm{MAE}$.

To conclude, we have demonstrated resonant tunneling spectroscopy of individual sulphur vacancy (V$\textsubscript{S}$) defects in monolayer MoS$_2$. From a comprehensive analysis of measured and DFT calculated in-gap states and defect wavefunctions, we have unravelled the respective roles of symmetry-lowering by a Jahn-Teller lattice distortion and strong ($\sim$100 meV) spin-orbit coupling. We have shown that the alignment of an unpaired spin-$1/2$ magnetic moment within the crystal plane is susceptible to local strain as detected in reduced defect wavefunction symmetry and orientation of their mirror axis. Future work will be directed towards investigating the possible role many-body effects in the electronic excitation spectrum \cite{Schulz2015}, as well as charge \cite{Rashidi2016} and spin \cite{Paul2017} dynamics.

\section{Experimental Section}
\textbf{Sample preparation:} The MoS$_2$ monolayers were mechanically exfoliated from a bulk geological crystal, and transferred using polydimethyl siloxane (PDMS) \cite{jain2018minimizing} onto epitaxial G/SiC prepared by flash annealing highly doped 6H-SiC substrate at 2000$^\circ$C in ultra-high vacuum (UHV) conditions. 1L-MoS$_2$ was identified by a combination of optical contrast and Raman/Photoluminescence spectroscopy (Supplementary Fig.1). Gold markers were deposited through a shadow mask onto the G/SiC substrate after preparation, but prior to the transfer of MoS$_2$ to help align the STM tip to the micron sized MoS$_2$ crystal. Before the STM measurements, the sample was annealed at 250$^\circ$C for 12 hrs in UHV conditions. 

\textbf{STM Measurements:} Low-temperature scanning tunnelling microscopy and spectroscopy (STM/STS) was performed in an Omicron low-temperature STM ($\sim$4.5 K) under UHV conditions ($\sim5\times10^{-11}$~mbar). For all spectroscopy measurements, we used electrochemically etched tungsten tips calibrated against the Au(111) Shockley surface state. Unless stated otherwise, the spectroscopy measurements were carried out using standard lock-in techniques with a modulation amplitude $V_{\rm ac}=$20 mV and a modulation frequency of 730 Hz. LDOS maps were taken in constant height mode with the lock-in modulation switched on. The
data were taken at a constant tip height with a lock-in modulation voltage ($V_{\rm ac}=$40 mV). In all experiments the tip was biased and the sample was grounded as indicated in Figure~\ref{fig:fig1}a.

\textbf{DFT calculation:} First-principles calculations were performed with the projector augmented-wave (PAW) method implemented in the Vienna Ab-initio Simulation Package (VASP) \cite{blochl1994projector,kresse1996efficient}. The Perdew–Burke–Ernzerhof (PBE) exchange-correlation functional \cite{perdew1996generalized} was employed with standard PAW pseudopotentials containing six valence electrons for sulphur (3$s^2$3$p^4$) and fourteen valence electrons for molybdenum (4$s^2$4$p^6$5$s^1$4$d^5$) \cite{kresse1999ultrasoft}. The predicted lattice constant of the primitive cell was 3.18 Å, consistent with the experimental value of 3.17 Å \cite{petkov2002structure}. A 5$\times$5$\times$1 supercell was then constructed with multiples of the optimized primitive cell. To avoid the interaction between monolayers in the periodic images, a vacuum layer of $\sim$15 Å was added. The cut-off energy for the plane wave basis representing the electronic wave functions was 500 eV. A sulphur defect was created by removing one surface sulphur atom. An additional electron was introduced to consider the sulphur vacancy in the $-1$ charged state. A homogeneous background charge was assumed in order to neutralize the Coulomb divergence induced by the charge in the simulation cell. For geometry optimization, the convergence criteria were set to $10^{-5}$ eV in energy and 0.01 eV/Å in force, respectively. Constant-height STM images were simulated within the Tersoff–Hamann model \cite{tersoff1983theory,tersoff1985theory} with the tip placed $\sim$3 Å above the surface. For this, the Fermi level was chosen so that the charge state V$_S^{1-}$ is occupied and V$_S^{2-}$ is unoccupied, and no symmetry of the defect wave function was artificially enforced. SOC was included in all calculations unless noted otherwise.

\begin{suppinfo}
MoS$_2$ layer thickness, Coulomb blockade in resonant tunneling, Fourier analysis of charge states, calculation of orbital contributions of in-gap states and Supporting Figures 1-6 (PDF). 
\end{suppinfo}

\section{Author Information}
\subsection{Corresponding Author}
*Bent Weber (email: b.weber@ntu.edu.sg)

\subsection{Author Contributions}
TA, GS, YW performed the scanning tunnelling spectroscopy experiments. GS fabricated the device. HM, RK, RR, BQ and TS performed the theoretical calculations. TA, RK and BW analyzed the data. RR and TS supervised the theory work. BW conceived and coordinated the project. TA, RK, and BW wrote the manuscript with input from all authors.

\begin{acknowledgement}

This research is supported by the National Research Foundation (NRF) Singapore, under the Competitive Research Programme ``Towards On-Chip Topological Quantum Devices'' (NRF-CRP21-2018-0001), with further support from the Singapore Ministry of Education (MOE) Academic Research Fund Tier 3 grant (MOE2018-T3-1-002) ``Geometrical Quantum Materials''. IV, CPYW and KEJG acknowledge the support from the Agency for Science, Technology, and Research (A*STAR) (\#21709). BW acknowledges a Singapore National Research Foundation (NRF) Fellowship (NRF-NRFF2017-11).
\end{acknowledgement}



\bibliography{main}

\end{document}